\documentstyle[aas2pp4]{article}
 
\newcommand{\be}{\begin{equation}}
\newcommand{\ee}{\end{equation}}

\newcommand{\etal}{ et al.\ }
\newcommand{\vs}{{\em vs}}

\newcommand{\labtab}[1]{\label{tab:#1}}
\newcommand{\labfig}[1]{\label{fig:#1}}
\newcommand{\labsecn}[1]{\label{sec:#1}}

\newcommand{\fig}[1]{Figure~\ref{fig:#1}}

\begin{document}
\title{Star Formation in the Interacting Pair NGC7733/34}
\author{ M. Jahan-Miri \altaffilmark{1} and H. G. 
Khosroshahi\altaffilmark{2} 
\affil{$^1$ Physics Department, Shiraz University, Shiraz 71454, Iran\\} 
\affil{$^2$Institute for Advanced Studies in
Basic Sciences, Zanjan, 45195, Iran\\} 
}
\altaffiltext{1}{jahan@physics.susc.ac.ir}
\altaffiltext{2}{khosro@iasbs.ac.ir}

\begin{abstract}
The problem of star formation within the interacting pair NGC7733/34
has been studied, based on the UBVRI photometry of the source. The
distribution of the colors of selected regions within the galaxies
is used to infer an estimate for the age distribution of the star
forming regions. The results seem to indicate the presence of numerous
extended young star-forming regions in the disk of one of the two galaxies, 
NGC 7733, with ages in the range of $10^6$--$10^8$~yr. However, there
exist no evidence for any violent star formation activity, in the
past $10^8$~yr, in the nuclei of the two galaxies. The pair seems to be 
a merger bound system with the brightest, youngest, site of star forming 
activity lying at the disk interface.     
\end{abstract}

\section{Introduction}
Interactions and mergers of galaxies are commonly considered to
trigger star burst activity in the disks as well as nuclei of the
spiral members. The occurrence of the activity in the different regions, and
in the different members, depends however on the associated time scales,
the geometry of the interaction, and the mass ratio of the member
galaxies (Toomre \& Toomre 1972). 
The pioneering work of Larson \& Tinsley (1978) examines the effects of
interaction on the colors of pairs. Normal Hubble galaxies have a well
defined narrow distribution on the [(U-B) \vs~(B-V)] diagram. Large scatter
in the two color diagram for the Arp Atlas against the well defined relation
in the same diagram for the Hubble Atlas (excluding Arp galaxies) is
understood as a result of the star formation histories of peculiar galaxies.
The effects of galaxy interactions in triggering star burst activity has
been well demonstrated through observational studies of the optical
(Kennicutt \etal 1987, Bushouse 1987), infrared (Lansdale \etal 1984), and
radio (Hummel 1981) emission of the interacting systems.

Interaction may lead to a high star formation rate in two different ways,  
ie. by increasing the formation rate per unit mass, and by increasing the
concentration of the gas in some regions of the galaxies (eg. Combes 1993).
The efficiency of the star formation process, per unit mass, has been
found to be higher, by a factor of 10, in the interacting galaxies than
its average value in the normal galaxies (Young 1993). Also, simulation
of the interacting galaxies has shown that the interaction will result in a 
redistribution of substantial quantities of the material into the central
regions of the galaxies. This injection of the fresh material (neutral
hydrogen) into the potential well of a galaxy is then expected to be followed
by a compression mechanism which could result in a rapid and efficient
formation of $H_2$ molecular clouds and their subsequent collapse to form
stars. The time scale for the nucleus to respond to the tidal disturbances
during the course of the interaction is however much less, by two orders of
magnitudes, than that for the disk (Joseph \& Wright 1985; Keel \etal 1985).
Hence, an observation of the starburst activity in the nucleus alone
would indicate that the interaction has started recently. In the more
probable case of a relatively old interaction history of the pair,
starburst is observed to be occurring contemporaneously throughout the two
member galaxies, both within and outside their nuclei (Alonso-Herrero
\etal 1999; Frayer \etal 1999; Vigroux \etal 1996). In contrast, the
interacting pair studied here might be an example of a rare case showing
star burst activity {\em only} within the disk of {\em only} one of the
two interacting spiral galaxies.

Here, we use UBV color mapping of the source NGC7733/34 to identify the
sites of violent star formation. The colors are then used to estimate the
ages of the selected young star forming regions. The implications of
the inferred age distribution, in contrast with the observed spatial
distribution of the regions in the different parts of the two galaxies,
is discussed, indicating a possible geometry for the interaction.
A more detailed and conclusive study of the source would nevertheless
require further spectroscopic data as well as radio mapping which we 
intend to carry out subsequently. 

\section{The source and the data}
\labsecn{data}
The source NGC7733/34 has been cataloged as a prototype for the subcategory
of spiral-spiral interacting galaxy pairs (Arp \& Madore 1987). NGC 7733
(RA 23:42:33.0, Dec -65:57:23; J2000)
is a barred spiral with a total B magnitude of 14.49, axis ratio 
1.3/0.8 arcmin, and heliocentric radial velocity 10199 km s$^{-1}$. 
The brighter other component, NGC 7734 (RA 23:42:42.8, Dec -65:56:41),
is also a barred spiral with a  magnitude 13.95, axis ratio 1.4/1.2 arcmin,
and radial velocity 10574 km s$^{-1}$ (de Vaucouleurs \etal, 1991;
Paturel \etal 1989).

In an investigation of the southern interacting galaxies, the gas content
of the pair has been estimated to include $\lesssim 4.2\times 10^9 M_\odot$
atomic hydrogen, in addition to the molecular gas with a mass
$\lesssim 15.4\times 10^8\,M_\odot$ in NGC7733, and
$\simeq 1.073 \times 10^{10} M_\odot$ in NGC7734 (Horellou \& Booth 1997).
Multi-aperture photo-electric photometry (in B \& V) of the  source among
several other southern spiral galaxies revealed that NGC7734 has a redder
B-V color at all scales, in the range 0.68--1.04 arcmin (Peterson 1986).
UBVRI as well as JHK photometry of the pair, among 59 other interacting
galaxies, has been reported by Johansson \& Bergvall (1990) giving diameters,
V magnitudes, and colors for the sample. Broadband images and spectra of
the pair may be found in Bergvall \& Johansson (1995). The source was
included in a study of the distribution in $W[NaI] - W[MgI]$ plane, which
is an indicator of the amount of the reddening of the starlight in the host
galaxy (Bica \etal 1991).

For the propose of this work, the images of the galaxy pair in five bands,
UBVRI, were obtained with the CCD Camera employed at the prime-focus
of the Anglo-Australian Telescope. Each CCD frame consisted of
512$\times$320 pixels with a pixel scale of 0.47 arcsec corresponding to
a spatial extent of $\sim 235$~pc, at the estimated distance of
$\sim 130$~Mpc to the source ($H_0=80 {\rm km \ s}^{-1} \ {\rm Mpc}^{-1}$).
The data reduction was carried out following the
standard procedure, with the use of appropriate flat-filed frames,
fringe-frames, and a bias-frame. The frames were made to coincide in the
pixel coordinates with the help of stellar images, and the sky light was
corrected for by averaging over areas far from the galaxies. The seeing
(FWHM) was estimated to be 1.5 arcsec corresponding to 3.2 pixels, based
on the radial profile of the point sources in the field. The B-band contour
maps of the two galaxies are shown in 
\fig{fig1a} (NGC7733) and \fig{fig1b} (NGC7734), with axes marked in pixel
coordinates.  The suspected starburst regions have been selected and
marked with numbers, on the two figures. The criteria for the
selection of these areas were based on their anomalously high B
brightness in the extra nuclear regions of each galaxy. These high
intensity extended knots are anticipated to be the sites of extensive
recent star formation activities, possibly being caused by the gravitational
interaction of the pair. The nuclei of the galaxies are also included
in the following analysis, identified as ``7733'' and ``7734'' in the figures.

\section{Results and discussion}
In order to look for the evidence for recent star formation activities in
the pair NGC7733/34 and also to determine the relative ages of the bursts
of star formation in the suspect regions, their two-color UBV diagram has
been plotted in \fig{fig2}. The (U-B) and (B-V) colors in \fig{fig2} for the
selected knots, indicated in Figure 1, has been derived by integrating the
flux over the central $3\times3$ pixels area of each knot in each band.
This corresponds to the central point source for each knot, given the
seeing condition (see above). The curve in \fig{fig2} represents the
theoretically predicted colors
for clusters with different ages, from the work of Searle \etal (1973) for
which we can find a tabulation of the cluster ages as well. The fine
details of the curve worked out in the more recent theoretical models
(eg., Girardi \etal 1995) are not relevant for the purpose of making a rough
estimate of the ages attempted here. In \fig{fig2} we have also plotted the
lines parallel to the reddening vector, passing through some of the knots.
Each line is marked by an associated age corresponding to its point of
intersection with the curve, based on the tabulation of the ages in
Searle \etal (1973). Table 1 lists the colors of the various selected regions
in the pair, as well as their inferred ages. The (systematic) errors in the 
individual colors are estimated to be $\lesssim 0.03$~ mag, based on the errors 
in the CCD calibration equations and the flatness of the flat field images 
used. The ages derived for the
starburst regions within the disturbed galaxy NGC7733 do not show any
specific order with respect to their relative spatial positions. This is
what one expects on theoretical grounds and had been also found in other
observational studies (Schweizer 1976; Jensen \etal 1981), albeit
the earlier expectations to the opposite (Larson 1978). The 
only definite conclusion in this regard is that the youngest star forming
regions, ie. knots~7, 6, \& 5 are located near the interface between
the two galaxies, as has been reported in other cases too
(Alonso-Herrero \etal 1999). Also, the longest inferred ages for the
starburst regions, hence the time since the start of the interaction, is
$\sim 10^8$~yr. This, together with the presence of a ``bridge'' between
the two galaxies, which is best seen in the I-band image (see, also, Arp
\& Madore 1987), is in accord with the theoretical expectations for the
formation of a bridge (Toomre \& Toomre 1972; Joseph \& Wright 1985) 
that require time scales  similar to the above value. 
Furthermore, the sites of recent activity, in NGC7733, are observed
to be located along a spiral pattern, as expected from the 
results of numerical studies of galaxy encounters. The formation
of central tidal arms, that move outward with time and later
evolve into density waves, originally seen in the two dimensional simulations 
that were restricted to symmetric disturbances (Toomre \& Toomre 1972) 
has been also verified by the three dimensional results which allow for 
non-symmetric perturbations as well
(Byrd \& Howard 1992). Moreover, it is seen in these studies that
the edge of the disk near the perturber is strongly disturbed forming
a separate tidal bridge arm, apart from the central arms (Byrd 1995). 
The ages and morphologies of
these arms may be in principle used to fix the parameters of an encounter,
via detailed simulations, as has been attempted in the case of
M51 (Byrd 1995).                                                          

The colors in \fig{fig2} have not been corrected for any internal reddening
due to the foreground gas and dust within the host galaxies.
The fact that all the knots lie to the right of the theoretical curve is
expected and may be simply explained in terms of the reddening effect of
the dust. Thus the true position of each knot on the curve could be
determined, by its displacement parallel, and opposite, to the reddening
vector which is along the lines indicated in \fig{fig2}. We have further tried 
to correct for the effect of the ``red'' population of old
foreground/background stars, based on the present broadband intensities  
which could not be however conclusive. This was done by choosing baseline
isophot contours around each knot, which correspond, in some cases, to
very large complexes of possibly correlated and clustered star forming
regions. The intensity, in each band, for the pixels
immediately surrounding the selected baseline was used to determine a
mean ``background'' contribution. This was then weighted by the area
of the interior region and subtracted from its integrated intensity, in
order to obtain its corrected colors. The resulting corrections were found
to be sensitive to the baseline selected, as would be expected. However,
one might argue that for the different, successively larger, baselines
all lying well within a true region of recent coherent star formation the 
corresponding corrected colors, for a given knot, would cause its position
on the UBV diagram to move systematically closer to the theoretical curve.
Thus, one might try to set an upper limit on the size of an
active region by finding the largest baseline for which the above
trend would persist. A baseline larger than the true size of the region
could be identified as it would result in a displacement of the position
of the knot on the diagram in a random direction. From this exercise, we find
the limiting size of the active regions to be as large as $\sim 1 kpc$,
in the case of knots 4, 7, and 9. This is further supported by the fact
that the uncorrected colors of these knots remains the same for any
selected area smaller than that limit, around the center of the knot.
In contrast, for the other knots the uncorrected colors vary greatly
depending on the size of the integration region. One may note that the
above method of using successive isophot baselines resemble that of 
multi-aperture photometry which has also been used in order to set limit
on the size of starburst regions (Campbell \& Terlevich 1984). 

A clear distinction between the star formation histories of the two
member galaxies may be seen immediately from the distribution of the
colors in \fig{fig2}. The colors of the starburst knots in NGC7733 indicate
recent activity, mostly with ages $\lesssim 10^7$~yr, being probably induced
by the gravitational interaction. However, the selected clusters
in NGC7734 are generally much older, having red colors around the red end
of the evolutionary track, also in agreement with the earlier observations
of the source (Peterson 1986). Moreover, the colors of NGC7734 lie, in
general, much closer to the curve, that is they have been subject to 
less reddening by the dust in the host galaxy, in comparison to those
in NGC7733. The scarcity of the dust in NGC7734 would in
turn account for its weak star formation activity, in contrast to the case
of NGC7733 which shows widespread activity as well as a rich supply of the
dust that causes the colors of the star forming regions to be shifted
further away from the curve. This conclusion seems however to be in
contradiction with the earlier estimates of the molecular gas content of
the two galaxies (Horellou \& Booth 1997). Alternatively, the lack of 
dust-reddening effects in NGC7734 may be attributed to an unfavorable 
geometry for absorption (see, eg., Thronson et~al. 1990), thus allowing for
the large reported molecular mass, as well.                     

The absence of star forming activity in NGC7734 is further confirmed from
the red colors of the selected knots in that galaxy
on the [(V-I) \vs~(V-R)] diagram, shown in \fig{fig3}. As is well
known, these colors are more sensitive, than UBV colors, to the presence
or absence of the underlying old stars (eg., Thuan 1985). Thus the very
red colors of the clusters of NGC7734, in Fig.~3, indicate even more
conclusively the
absence of young population in that galaxy. One may, in particular, note 
from Fig.~3 that knot~29 is indeed populated only by old stars, even though
its UBV colors in \fig{fig2} might erroneously imply it to be the site of 
recent star formation. 
The blue UBV colors might, as well, be caused by the selective
obscuration (the ``picket-fence'' effect) of a foreground clump
with different covering factors for the gas and dust (Calzetti 1997).  
Therefore, gravitational
interaction of the two galaxies has not induced any burst of star formation
activity in the more massive party, ie. NGC7734, as has been reported
in other cases (Bushouse \& Gallagher 1984). Theoretical simulations of
pair interactions also indicate cases where the more massive party is not
much affected (Toomre \& Toomre 1972). That NGC7734 is the more massive
member in the pair is evidenced by its larger extent as well as its larger
luminosity (de Vaucouleurs G., \etal, 1991; Paturel \etal 1991).
The geometry of an encounter as well as the timescales for the 
response of each member galaxy are also among the other factors that determine 
which member is influenced more by the interaction. The redistribution
of the gas in a disk can be very different depending on the relative
directions of the disk rotation and the encounter orbit, as is evidenced by the 
morphologies of the observed pairs, and is  also verified by the results of n-body 
simulations.          
However, and somewhat paradoxially, the star formation rates
do not seem to obey a close correspondence with the morphological
changes induced by encounters (Keel 1993). The star formation
rates are often found to be greatly enhanced in only one member of
the pairs, based on the studies of far-infrared emissions in the
IRAS
survey (Telesco \& Wolstencroft 1988) and  the UV-bright Markarian
galaxies (Keel \& van Soest 1992); a result which makes sense also for
the statistics of the velocities versus star formation rates of the pairs in 
the CfA redshift survey
(Barton \etal, 2000). The discrepancy might be, in part, caused by the
theoretically expected time delay between  the greatest gravitational
perturbation and the triggering of the star formation in the interstellar
medium.                                                                   

\fig{fig2} gives another important clue for the effects of the interaction
on the star formation activity in the galaxies. The colors of the central
regions of both galaxies are seen to be located at the extreme red end of
the distribution. The two nuclei have actually the reddest colors, 
$(B-V) \sim 1$, among all the selected regions. This indicates the absence of 
nuclear starburst activity in the pair, which is further supported by
the very low infrared-luminosity of the pair (Persson 1988).
The absence of nuclear activity, among interacting galaxies, have been
reported in other sources as well (Bushouse \& Gallagher 1984). If the
upper limit on the age of starburst regions in NGC~7733 ($\sim 10^8$~yr)
is taken as the time since the interaction has triggered the activity in
the pair, a quiet nucleus seems an odd finding. Given the much shorter
timescale required for the nuclear response to a tidal disturbance, the
star formation in the nucleus of the galaxy must have started long ago
but did not continue vigorously. An ongoing nuclear starburst requires
the supply of fresh material from the disk or the companion (Joseph \&
Wright 1985) which must have failed in the case of NGC7733/34. 

The red colors of the nuclear region of NGC7733 might have been affected
also by the dust lying in that direction. A large column density for the
dust in the line of sight to the center of NGC7733 is also in accord with
its observed major/minor axes ratio which imply a relatively tilted geometry.
Also, inspection of the radial profiles of the colors (not presented here)
along two radial directions one passing through knots 2 \& 3 and the other
through knot 8 indicates the presence of dust lanes along those radial
directions, which could contribute to the obscuration of the center of the
galaxy as well. Nevertheless, nuclear starburst activity as is observed in
many interacting galaxies (eg., Alonso-Herrero et~al. 1999) is ruled out
for NGC7733, because of the very weak IR-luminosity of the source. 
Therefore, the pair studied in this work provides a case where recent
starburst activity is observed only within the disk of only one of the
two interacting spiral galaxies. However, a more comprehensive study
based on additional spectroscopic and radio data is needed in order to
quantify the above conclusion and be able to look for its justification.

Finally, it may be noted that the youngest and the brightest star forming
complex in NGC7733, that is knot 7, is located near the apparent disk
interface, a property that has been seen in other cases too, eg. NGC4038/9, and
NGC6621/2. Given the inferred age of knot 7, $\sim 10^6$~yr, and assuming 
its activity was triggered when it was located at the "retarded" interface 
of the two galaxies, a geometry of the orbit that would imply its apparent 
displacement ($\sim 24$~Kpc) from the line joining
the centers of the galaxies to be caused by its orbital motion is
ruled out. Because, the corresponding velocity would otherwise amount to 
$\sim 20,000$~km/s, which is in contradiction with the expected bound 
nature of the orbit as
implied by the inferred $10^8$~yr old interaction history of the pair.
The two galaxies could had already formed a bound system before
their most recent interaction, which nevertheless makes the orbit tighter, on the
statistical grounds that there exist too many close pairs to be accounted 
for by chance encounters of initially distinct galaxies (Chatterjee 1987).
However,
knot 7 may as well be presently located at the true interface between
the two disks, noticing that the planes of the two galaxies are tilted in 
different directions and at different angels, as implied by their aspect ratios.

\acknowledgements
We are grateful to the Anglo-Australian Observatory for the observations.
MJ is thankful to AAO also for the use of their computer and image
processing facilities while carrying out the data reduction. We wish to thank 
the referee for making several useful suggestions/corrections that helped 
to improve the presentation. This research (MJ) was supported by a grant from 
the Research Committee of Shiraz University.

\begin{deluxetable}{clllll}
\tablewidth{0pt}
\tablecaption{Colors and Ages of the selected $knots$}
\tablehead{\colhead{Knot}&\colhead{$B-V$}&\colhead{$U-B$}&\colhead{$V-R$}&\colhead{$V-I$}&\colhead{$Age$ (yr)}}
\startdata

  1  &  0.31 &  -0.54 &  0.36 & 0.52 & $5\times 10^6$\\
  2  &  0.27 &  -0.47 &  0.26 & 0.56 & $10^7$\\
  3  &  0.35 &  -0.33 &  0.35 & 0.71 & \\
  4  &  0.32 &  -0.32 &  0.41 & 0.62 & \\
  5  &  0.42 &  -0.57 &  0.33 & 0.62 & \\
  5' &  0.35 &  -0.55 &  0.38 & 0.64 & $3\times 10^6$\\
  6  &  0.24 &  -0.57 &  0.24 & 0.45 & \\
  6' &  0.28 &  -0.44 &  0.47 & 0.90 & $10^7$\\
  7  &  0.27 &  -0.72 &  0.20 & 0.39 & $10^6$\\
  8  &  0.63 &  -0.02 &  0.43 & 0.84 &\\
  9  &  0.32 &  -0.56 &  0.39 & 0.52 & $3\times 10^6$\\
  10 &  0.31 &  -0.22 &  0.37 & 0.69 & \\
  11 &  0.21 &  -0.28 &  0.36 & 0.55 & \\
  12 &  0.52 &  -0.47 &  0.30 & 0.49 & \\
  20 &  0.63 &  ~0.22 &  0.43 & 0.95 & $3\times 10^8$\\
  21 &  0.92 &  ~0.43 &  0.53 & 1.13 & \\
  22 &  0.70 &  ~0.19 &  0.60 & 1.19 & \\
  23 &  0.68 &  -0.04 &  0.54 & 1.12 & $5\times 10^7$\\
  24 &  0.51 &  -0.20 &  0.52 & 0.88 & \\
  25 &  0.74 &  ~0.19 &  0.52 & 1.02 & $10^8$\\
  26 &  0.66 &  ~0.17 &  0.60 & 1.16 & \\
  27 &  0.52 &  -0.03 &  0.41 & 0.88 & $10^8$\\
  28 &  0.52 &  ~0.09 &  0.45 & 0.94 & \\
  29 &  0.50 &  -0.63 &  0.54 & 0.95 & \\
  19 &  0.70 &  ~0.22 &  0.54 & 1.15 & \\
7734 &  1.00 &  ~0.66 &  0.67 & 1.24 & \\
7733 &  0.99 &  ~0.47 &  0.63 & 1.18 & \\
\enddata
\labtab{colors}
\end{deluxetable}

\newcounter{f1}
\setcounter{f1}{1}
\newcounter{f2}
\setcounter{f2}{1}
\renewcommand{\thefigure}{\arabic{f1}\alph{f2}}
\begin{figure}
\epsscale{1.0}
\plotone{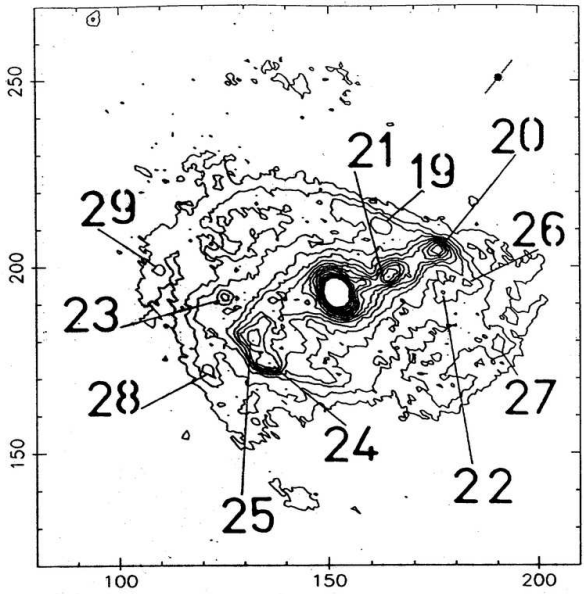}
\caption{Isophot contour plot of NGC7733 in B band; axes are marked by
the absolute pixel coordinates on the CCD. The suspected starburst
regions are marked by numbers. The bar at the {\em top right} indicates
the relative direction to the companion galaxy, while the {\em dot}
on it identifies a fixed location for comparison with Fig.~1b. The relative 
position of the pair may be also reckoned from the pixel coordinates.}
\labfig{fig1a}
\end{figure}

\setcounter{f2}{2}
\begin{figure}
\epsscale{1.0}
\plotone{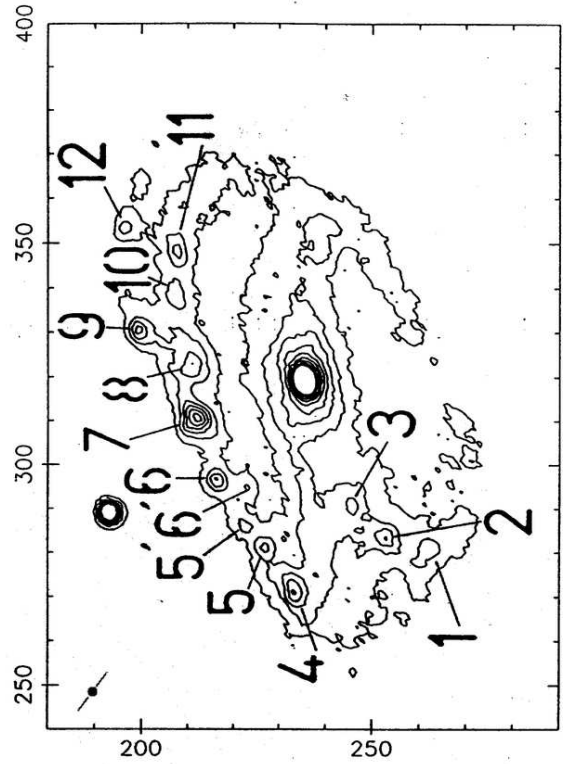}
\caption{Same as Fig.~1a, but for the other member of the pair, NGC7734.
The bar at the {\em bottom left} serves for the similar purpose as in
Fig.~1a}
\labfig{fig1b}
\end{figure}

\renewcommand{\thefigure}{\arabic{figure}}
\setcounter{figure}{1}
\begin{figure}
\epsscale{1.0}
\plotone{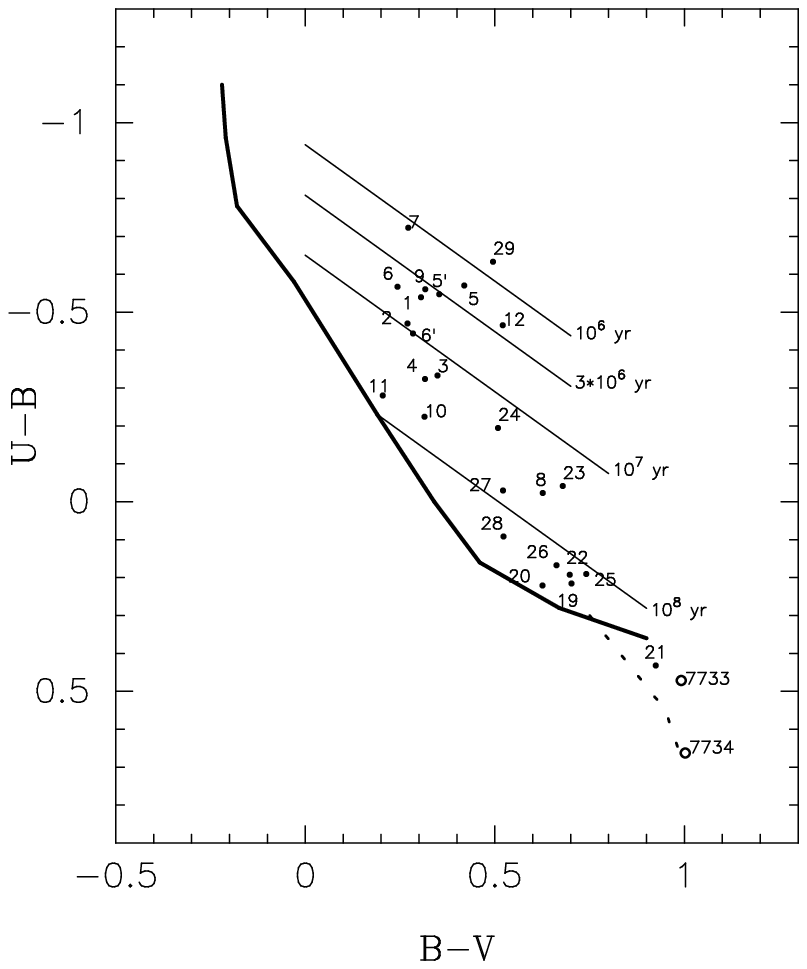}
\caption{The UBV two-color diagram for the selected regions, derived by
integrating intensity values over the 9 central pixels for each region. 
The colors have not been corrected for the reddening in the host galaxies.
The {\em thick} curve is the computed relation from Searle \etal (1973), 
and its continuation in the red ({\em dashed}) is taken from Jensen (1981).
The {\em thin} lines are drawn parallel to the reddening vector, each 
being marked with an age value corresponding to its crossing point with
the curve.}
\labfig{fig2}
\end{figure}

\begin{figure}
\epsscale{1.0}
\plotone{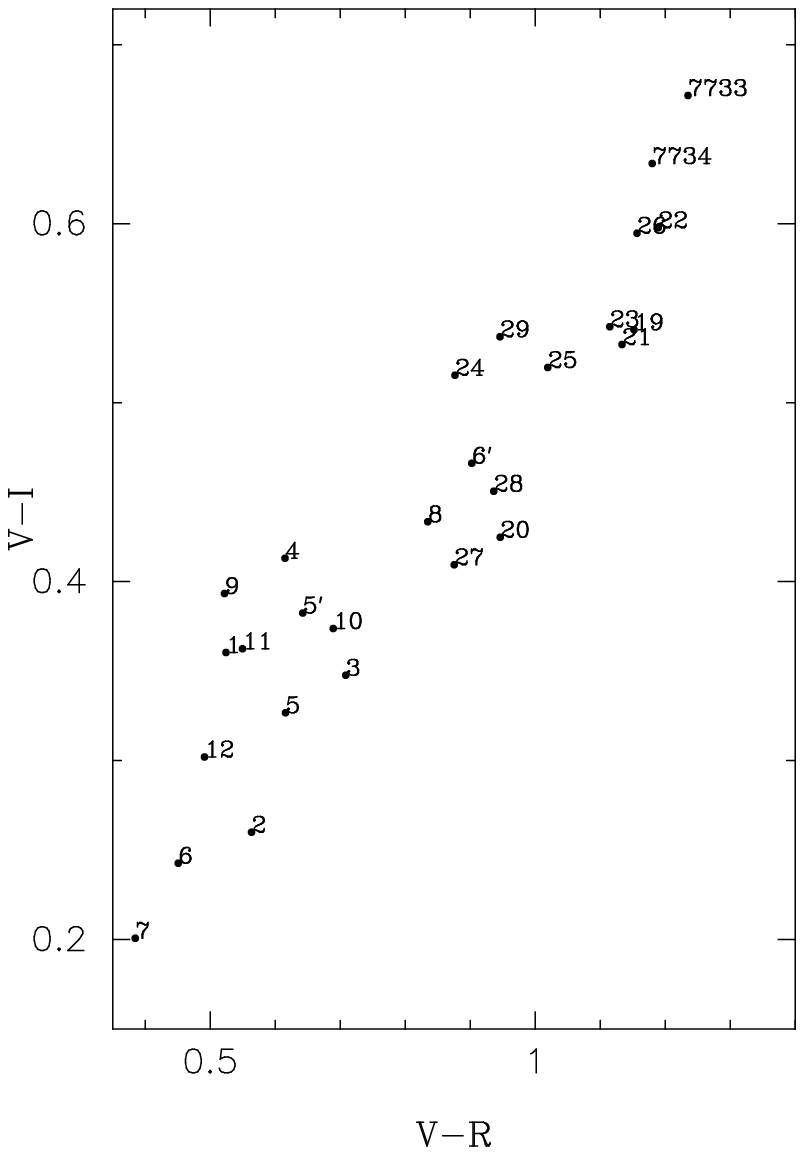}
\caption{The VIR two-color diagram for the selected regions in
the two galaxies, indicated by their numbers.}
\labfig{fig3}
\end{figure}

\end{document}